\documentclass[aps,pra,twocolumn,groupedaddress,amsmath]{revtex4-2}%
\usepackage[paperwidth=215.9mm,paperheight=279.4mm,centering,hmargin=1.8cm,vmargin=2cm]{geometry} %

\usepackage{graphicx}
\usepackage{dcolumn}
\usepackage{bm}
\usepackage{url}%

\begin{document}

\title{Fast qubits of optical frequencies on the rare-earth ions in fluoride crystals \\ 
and color centers in diamond}

\author{V.~Hizhnyakov$^{1}$}
\email[]{hizh@ut.ee}
\author{V.~Boltrushko$^{1}$}
\author{A.~Shelkan$^{1}$}
\author{Yu.~Orlovskii$^{1,2}$}
\affiliation{$^{1}$Institute of Physics, University of Tartu, W. Ostwald Street 1, 50411 Tartu, Estonia}
\affiliation{$^{2}$Prokhorov General Physics Institute, Russian Academy of Science, 38 Vavilov street, 
119991 Moscow, Russia}

\date{\today}

\begin{abstract}
Fluoride crystals doped with rare-earth ions (REI) and pair centers in diamond for fast  ($10^{-9}\rm{s}$)
quantum computers (FQC) are proposed. As specific systems for REI doping, we propose
$Ca_{1-x}Sr_{x}F_{2}$ crystals and their analogues.
The $^{4}f$-states with different total orbital angular moments of these ions serve as
two-level systems (qubits). Suitable REIs are proposed as well. It is established that the pair $SiV$
and $GeV$ centers in diamond are promising systems for fast optical quantum computers operating
at elevated temperatures.
\end{abstract}

\maketitle


\section{Introduction}%

We discuss the possibility of creation fast (GHz) optical quantum computers (OQCs)
using mixed nanocrystals doped with rare-earth ions (REI) and diamond with pair $SiV$ and $GeV$
optical centers (OCs).
The qubits in these systems have optical frequencies and can operate at high (GHz) speed.

Recently we have found \cite{ref-hizhnyakov21} that crystals  $Ca_{1-x}Sr_{x}F_{2}$ and their analogs doped with REIs
can be considered as promising candidates for the role of fast OQC. In the specified
crystals one must use REIs that have $^{4}f$ states with small
and large diagonal elements of the Judd-Ofelt matrix $U^{(2)}$; the states with a small elements can
be used for qubits, and the states with large elements can be used for
implementing CNOT and other conditional operations.

In a number of previous studies of the use of crystals doped with REIs as quantum
computers, the authors considered the use of hyperfine levels as qubits
(see \cite{ref-wesenberg07} and the references therein). One-qubit operations with
such qubits require two optical $\pi$-pulses. These pulses must have a small
spectral width, much less than the qubit frequency. Therefore, such qubits
can operate with millisecond or longer sampling time. 

However, if to use REIs qubits with optical frequencies, then the spectral width of the light pulses 
will be incomparably less than the frequencies of the qubits. In this case, the light pulses will 
be much shorter than in the case of qubits with hyperfine levels. 
The weak interaction of $^{4}f$-electrons with surroundings
allows to have qubits with a long coherence time. In this case one-qubit gate operations
can be performed using single light pulses; conditional gate operations can
be performed using multiple optical pulses and a Stark blockade \cite{ref-hizhnyakov21}.

This article also draws attention to the fact that the pair centers of REIs and of other ioins 
can also be used as fast optical frequency qubits for OQCs.  Here we take into
account that such centers have cooperative excited states: dark and bright. Dark
states can be used for qubits, and bright states are suitable for implementing a conditional
gate operations. Pair $ SiV $ and $ GeV $ centers in diamond are especially promising 
for the development of fast OQCs. The advantage of such a OQC is 
that it can operate at elevated (possibly nitrogen) temperature.

\section{Materials and Metods}

\subsection{Mixed fluoride crystals doped with REIs}

The mixed $Ca_{1-x}Sr_{x}F_{2}$  crystals doped
with REI has a large inhomogeneous width $\Gamma_{inh} \sim$ THz and small homogeneous 
width $\Gamma_h \lesssim$ MHz of zero
phonon lines (ZPLs) in their optical spectra. This is of great importance for fast
OQCs, since a short sampling time $\tau \sim$ ns
requires the use of laser pulses with a spectral width $\Gamma_{L} \sim$ GHz.
The condition $\Gamma_{inh} \gg \Gamma_{L}$ ensures that a
large number of qubits can be addressed individually using laser pulses with
different frequencies.

To perform the CNOT gate operation, it is necessary to violate the
excitation of the target qubit by changing of the state of the control
qubit. If the used states interact so strongly that the frequency shift
$\delta$ at this change exceeds $\Gamma_{L}$ then this condition is
fulfilled. Previously the exchange dipole-dipole interaction was considered as the
main interaction responsible for this phenomenon. Therefore, the term
"dipole blockade" was used to denote it \cite{ref-wesenberg07}. 
However, we have found that, due to
small oscillator strength of $^{4}f-^{4}f$
transitions, for small and intermediate distances the strongest interaction
is the static (Stark) quadrupole-quadrupole interaction (see \cite{ref-hizhnyakov21}).
Therefore, here we use the term Stark blockade. We also found that the Stark 
blockade is determined by the diagonal matrix elements of the Judd-Ofelt matrix $U^{(2)}$.
These matrix elements have extremely scattered values, which differ by many
orders of magnitude. This allows to use the $^{4}f$ levels with
small diagonal elements of the matrix $U^{(2)}$ as qubit levels
$\lvert{0}\rangle$ and $\lvert{1}\rangle$; and the
$^{4}f$ levels with large diagonal elements of the matrix
$U^{(2)}$ as auxiliary levels $\lvert{1'}\rangle$ to implement
conditional gate operations \cite{ref-hizhnyakov21}.

\subsection{Pair centers in diamond and fluoride crystals}

Pair REI centers  in fluoride crystals and others can also be used as fast optical frequency
qubits for OQCs. This is due to the fact that such centers have cooperative excited states: 
dark and bright. Dark states can be used for the qubits themselves, and bright states are 
suitable for conditional gate operations. Paired $SiV$ and $GeV$ centers in diamond are especially 
promising for fast OQCs, because, due to the very high Debye frequency, they can operate 
at elevated temperatures.

\subsection{Laser excitation}

The OQCss with electronic states of impurity centers in crystals as qubits can operate using
a sequence of resonant light $\pi$-pulses. The intensity of such a pulse is \cite{ref-hizhnyakov21}
\begin{equation}
I=4 \pi^2 \hbar \Gamma_L^2 k^3/3 \gamma_0 Z, \nonumber
\end{equation}
where $k=\omega n/c$ is the wave number of light with frequency $\omega$, $n$ is refractive index, 
$Z=376.7$ ohm is the impendance of free space. For $k=20000$ cm$^{-1}$, $\gamma_0 =1.5$ ns and 
$\Gamma_L = 10^9$ sec$^{-1}$ the intensity of the ligth of the pulse is $I \sim 10^2$ W/cm$^2$.
The energy of the pulse E$_L=IS/\Gamma_L$ for the beam crossection $S=10^{-7}$ cm$^2$  is  10$^{-14}$ J. 
For REIs as qubits, the pulse energy should be about four orders of magnitude higher due to the rather 
low rate of radiative decay  $\gamma_0$ \cite{ref-hizhnyakov21}.
The corresponding field strength $\sim 3\cdot 10^{4}\,\rm{V/cm}$, although high, is still four  
orders of magnitude less than the interatomic field strength, which means that it does not lead to 
radiation damage to the crystal.

\section{Results}

\subsection{Rare earth ions and their $^4f$ -states suitable for fast OQCs}%

\begin{figure}
\centerline{\includegraphics[width=9 cm]{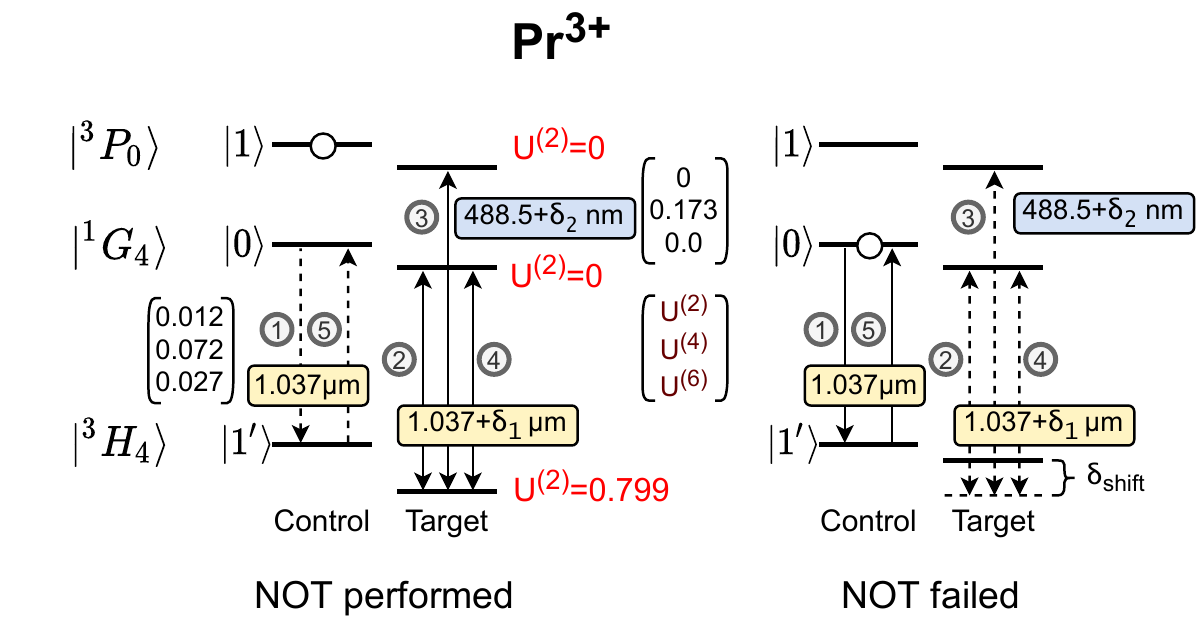}}
\caption{CNOT operation for $Pr^{3+}$ ions.}
\label{fig:pr3}
\end{figure}

For OQC, $Pr^{3+}$ ions can be used with the ground state $^{3}H_{4}$ as the auxiliary states
and the states $^{1}G_{4}$ (in $Pr^{3+}:YLiF_{4}$, $\tau=14\mu$s, see \cite{ref-basiev96})
and $^{3}P_{0}$ (in $Pr^{3+}:LaF_{3}$, $\tau=55\mu$s, see \cite{ref-brown65})
as qubit states (Figure~\ref{fig:pr3}).
Optical transition between all these levels is reasonably
allowed. The calculations begin with preparation of states $\vert 0\rangle $
using the $\pi$-pulse excitation of the 
$\left|^{3}H_{4}\right\rangle \rightarrow \left|^{1}G_{4}\right\rangle $ transition.

In figures 1-3, one can see the scheme of CNOT operation for the $Pr^{3+}$, $Er^{3+}$, and $Tm^{3+}$ ions
in mixed doped microcrystals with large inhomogeneous broadening.
The arrows indicate the excitation by $\pi$- light pulses, the numbers denote
the pulse sequence. The circles indicate the initially occupied levels of the control qubits.
${\delta}_{\rm{shift}}$ is a shift of the auxiliary level in the target qubit by changing the state
of the control ones. ${\delta}_{i}$ is a shift of the energy of the target optical qubit level
relative to the control one. The squares of the reduced matrix elements $U^{(k)}$ of the inter-level
transitions are in brackets. The intra-level $U^{(k)}$ values are near the corresponding levels.

\begin{figure}
\centerline{\includegraphics[width=9 cm]{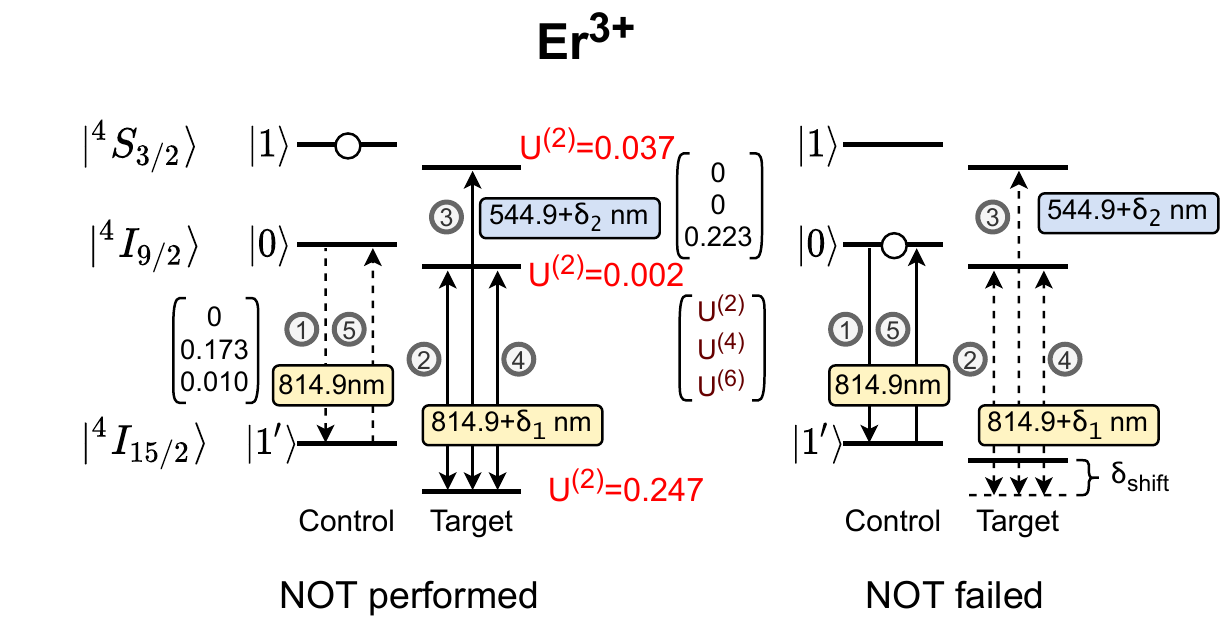}}
\caption{CNOT operation for $Er^{3+}$ ions.}
\label{fig:er3}
\end{figure}

\begin{figure}
\centerline{\includegraphics[width=9 cm]{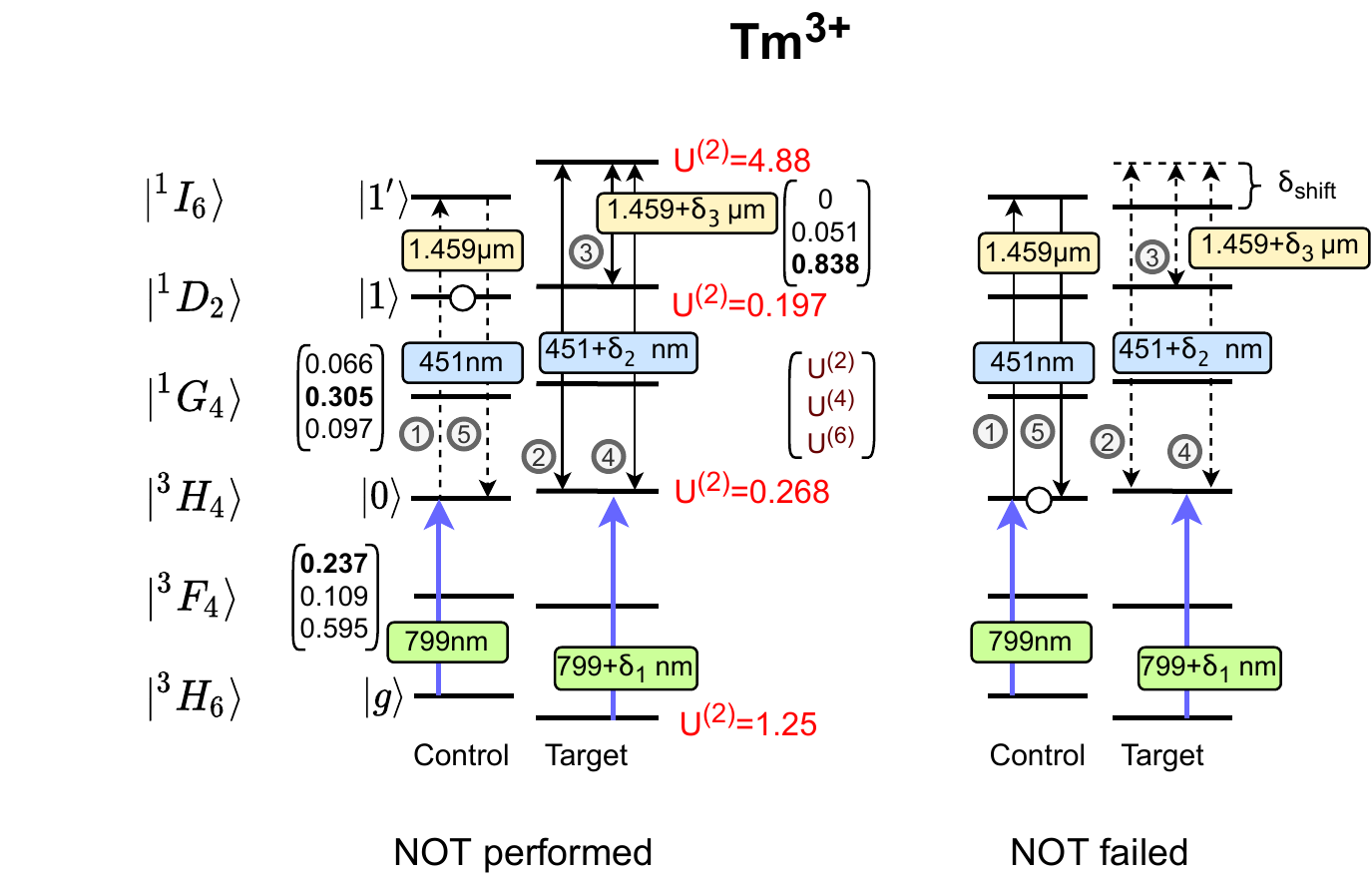}}
\caption{CNOT operation for $Tm^{3+}$ ions.}
\label{fig:tm3}
\end{figure}

The relative position of the levels can be different.

The $Er^{3+}$ ion ground state $^{4}I_{15/2}$ suits for 
the auxiliary state and states $^{4}I_{9/2}$ (in $Er^{3+}:LaF_{3}$, $\tau =133\mu$s, see \cite{ref-wesenberg07})
and $^{4}S_{3/2}$ (in $Er^{3+}:LaF_{3}$, $\tau =923\mu$s, \cite{ref-okamoto75}) as qubit levels.

The $Tm^{3+}$ ion provides several choices for the qubit and 
auxiliary level. For example, the lowest level $\left| ^{3}H_{6}\right\rangle $ of the
$Tm^{3+}$ ion has a rather large diagonal element of matrix
$U^{(2)} =1.25$ (Figure~\ref{fig:tm3}), and it can be used as an auxiliary level
$\left|1'\right\rangle $. In this case, levels $\left|0\right\rangle =\left|^{3}F_{4}\right\rangle $
($E=5619\,\rm{cm}^{-1}$, lifetime $\tau =18.05\,ms$ in $Tm^{3+}:LiYF_{4}$, \cite{ref-hizhnyakov21})
and $\left|1\right\rangle =\left|^{1}D_{2} \right\rangle $
($E=27830\,\rm{cm}^{-1}$, lifetime $\tau =70\mu\rm{s}$ in $Tm^{3+}:LiYF_{4}$, \cite{ref-walsh98}) have much smaller
diagonal element of matrix $U^{(2)}$ and can be used as qubit levels.
Another option (Figure~\ref{fig:tm3}) is that the initial state $\left|0\right\rangle $
is prepared by applying a pulse with frequency of $12518\,\rm{cm}^{-1}$
to the allowed transition $\left|g\right\rangle \to \left|0\right\rangle $
from the ground state $\left|g\right\rangle =\left|^{3}H_{6}\right\rangle $. In the
case of the $Tm^{3+}$ ion, high energy level $\left|^{1}I_{6}\right\rangle$
($E=34684\,\rm{cm}^{-1}$, lifetime $\tau =300\mu$s in
$Tm^{3+}:\beta-NaYF_{4}$, see \cite{ref-shi11}) has especially large
diagonal element of matrix $U^{(2)}$ ($\vert U_{^{1}I_{6}^{1}I_{6} }^{(2)}
\vert^{2}=4.88)$, and it can be used also as auxiliary level. Then one can
use the levels $\left|0\right\rangle =\left|^{3}H_{4}\right\rangle )$
and $\left|1\right\rangle =\left|^{1}D_{2}\right\rangle$ as qubit levels. Very large value of
$\vert U_{^{1}I_{6}^{1}I_{6} }^{(2)} \vert^{2}$ suggests that this scheme
may be used for implementation of CNOT gate in case of large mean distance
between $Tm^{3+}$ ions.

\subsection{Pair centers}

\subsubsection{Electronic states of pair OCs}

Among the very wide variety of crystals doped with various impurities, one can find many such crystals 
in the optical spectra of which there are strong zero-phonon lines with small homogeneous and large 
inhomogeneous width. Such crystals can be of interest as working elements of optical quantum computers, 
although individual centers may not have both strongly and weakly interacting states required for one-qubit 
and two-qubit gate operations. Indeed, this disadvantage of single centers can be overcome if
a pair of such centers located nearby is used for a qubit. It is taken into account that the paired center 
has cooperative excited states: dark and light. Dark states interact weakly and can be used for qubits. 
The dipole-dipole exchange interaction of bright states is strong (compared to dark states), so they 
are suitable for the implementation of conditional gate operations.

The energies and the wave functions of the stationary states of the pair OC consisting
of two almost identical centers are:
\begin{align*}
E_0& \simeq E-\Delta, &\; \left|0\right\rangle &\simeq 2^{-1/2}((1-\varepsilon/\Delta)\Psi_1(x_1)\Psi_2(x_2)\\
& & &-(1+\varepsilon/\Delta)\Psi_2(x_1)\Psi_1(x_2)), \\
E_1&=0, &\; \left|1\right\rangle &= \Psi_1(x_1)\Psi_1(x_2), \\
E_1'&\simeq E+\Delta, &\; \left|1'\right\rangle &\simeq 2^{-1/2}((1-\varepsilon/\Delta)\Psi_1(x_1)\Psi_2(x_2)\\
& & &+(1+\varepsilon/\Delta)\Psi_2(x_1)\Psi_1(x_2)), \\
E_2&=2E, &\; \left|2\right\rangle &= \Psi_2(x_1)\Psi_2(x_2).
\end{align*}

\begin{figure}[h]
\centerline{\includegraphics[width=7cm]{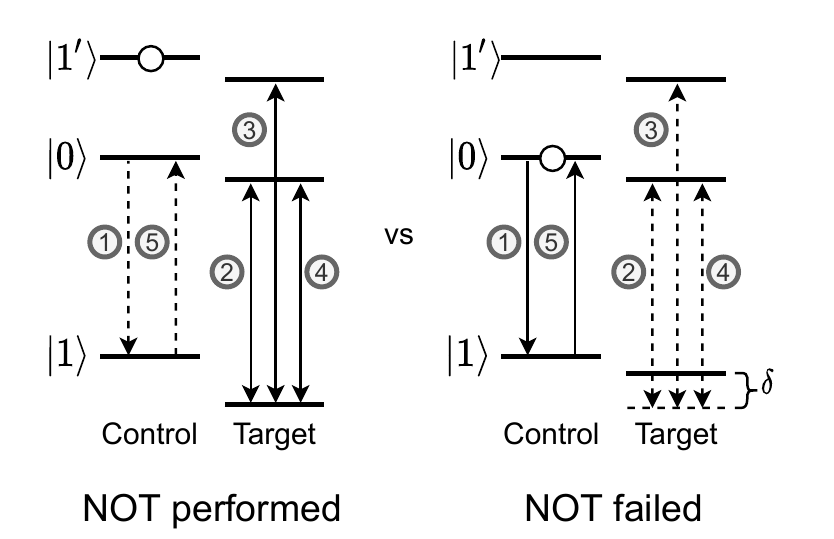}}
\caption{CNOT gate scheme for pair centers. The arrows indicate
$\pi $-pulses of light, the numbers next to them are the numbers of the
pulses in the sequence. The circles indicate the initially occupied levels. 
$\delta $ is the shift of the auxiliary level in the
target qubit due to a change of state of the control qubit.}
\label{fig:cnot}
\end{figure}

\noindent
Here $E_0 \pm \varepsilon$ are the differences in the energy of the excited ($\Psi_2$) and ground ($\Psi_1$)
states of the single centers forming the pair OC center, $\Delta$ is the exchange interaction in the pair OC 
assumed that half of the energy differenc $\varepsilon$ between the levels of the single centers forming 
a pair OC is very small as compared to the exchange interaction energy ($\varepsilon \ll \Delta$). 
In this case $\left|0\right\rangle$ is the metastable (dark) state, while
$\left|1'\right\rangle$ is the radiative (bright) state with $(\Delta/\varepsilon)^2$ times larger oscillator
strength of the dipole transition to the ground state $\left|1\right\rangle$ than of the dark state and twice 
large oscillator strength than the oscillator strength of the single center.
Due to the large oscillator strength, the exchange dipole-dipole interaction of two pair OCs in the states
$\left|1'\right\rangle$  is strong. Therefore, the dipole blockade in these states is also strong \cite{ref-lukin01}, 
and these states can be
used to implement the CNOT and othe conditional gate operations. On the contrary,
two pair OCs in a dark state $\left|0\right\rangle$ interact very weakly. Therefore, these states 
can be used as the states of qubits.

\subsubsection{$SiV$ and $GeV$ pair centers in diamond}

Especially promising are pair $SiV^{-}$ and $GeV^{-}$ centers in diamond due to large oscillator strength
of the optical transitions in the centers, high quantum yield of these transitions and very high Debye 
frequency of diamond allowing to work at elevated temperature.
Other important advantages of these centers are that,
due to very large oscillator strength of the optical transitions, it is possible
to use relatively weak laser pulses and crystals having a low concentration of OQs
$\sim$ 10$^{-4}$ -- 10$^{-6}$.

The single $SiV$ and  $GeV$  OCs have half a spin. Therefore, pair $SiV$ of $GeV$
OC can be in the singlet and triplet states. However, due to singlet-triplet relaxation, only excited states with lower
energies can be of interest. The energy level diagram of a pair OC consisting of two half-spin OCs is given 
in Figure 3 in \cite{ref-orlovskii20}.

A schematic diagram of the CNOT gate operation using paired centers is shown in Figure~\ref{fig:cnot}.

Note that recently pair $NV$ centerts in diamond, similar to pair $SiV$ and $GeV$ centers,
were investigated in \cite{ref-chou18}.
%

\subsubsection{Pair centers of rare earth ions: $Nd^{3+}$ in $CaF_2$}%

Pair optical centers of REIs are also promising systems for OQCs due to rather weak interaction
with an environment and a large number of electronic states in a center with different properties \cite{ref-hizhnyakov21}.
 which allow to fulfill corresponding criteria (DiVincenzo's criteria); see, for example \cite{ref-wesenberg07}.
Additional oportunities arise if we use the pair REI OCs.

Compared to the $SiV$ and $GeV$ OCs in diamond, the REI OCs have a disadvantage -- optical transitions in
these centers have a much lower oscillator strength and, therefore, these centers interact  much weaker with each other.
However, this weakening of the interaction can be compensated for by the possibility of using a high REIs concentration. 
Moreover, if doped crystals with divalent cations are used, e. g, fluoride type crystals  
($CdF_2$, $CaF_2$, $SrF_2$, $BaF_2$) then single
centers can exist only with interstitial  $F^{-}$ ions as charge compensators. Such centers have a large static 
dipole moment and interact quite strongly. This allows a large number of single centers to be converted into pair 
ones, for example,  using heat treatment.

To make a microcrystal to work as an OQC instance, we must find the frequencies of N $\sim 10^2$ closely
spaced pair OCs. This can be done, using the spectral hole burning and gain (anti-hole)
saturation method, see  \cite{ref-hizhnyakov21}.

\section{Discussion}

Here we indicated that optical centers of rare-earth ions (REIs) in fluoride crystals and paired centers 
in these crystals and in diamond can be used to create fast (with sampling time $10^{-9}\rm{s}$)  
optical quantum computers (OQC). As specific crystals for REI doping, we propose
$Ca_{1-x}Sr_{x}F_{2}$ crystals and their analogues.
The $^{4}f$-states with different total orbital angular moments of REIs in these crystals serve as
two-level systems (qubits). 

We have found that the $^{4}f$ levels with
small diagonal elements of the matrix $U^{(2)}$ can be used as qubit levels
$\lvert{0}\rangle$ and $\lvert{1}\rangle$; and the
$^{4}f$ levels with large diagonal elements of the matrix
$U^{(2)}$ can be used as auxiliary levels $\lvert{1'}\rangle$ to implement
conditional gate operations.
A few specific rare earth ions and the corresponding $^{4}f$ levels of these ions have been proposed 
for OQC (see Figs. 1-3). 

It was also found that, on the whole, crystals with strong ZPLs with a small homogeneous and large 
inhomogeneous width  in optical spectra may be of interest as possible candidates for qubits of OQCs, 
even if individual centers do not have both strongly and weakly interacting states necessary for 
one-qubit and two-qubit gate operations. The reason is that this disadvantage of such single centers 
can be overcome by using a pair of such centers located nearby for the qubits. It is taken into account 
that the paired centers have cooperative excited states: dark and bright. Dark
states can be used for qubits, and bright states are suitable for implementing a conditional 
gate operations.

It is also found that the $SiV$ and $GeV$ pair OCs in diamond and the REI pair OCs 
in fluoride-type crystals are promising systems for use as optical frequency qubits in fast OQSs.
In diamond with low concentration of $SiV$ and/or $GeV$ OCs, for the successful use of paired 
OCs for fast OCs, a significant part of the centers must be converted to paired  OCs. 
If all single centers will be converted to pair OCs
then the required minimum concentration will be  $c \sim (a/R_0)^3 \sim 10^{-5}$.

Pair OCs of $Nd^{3+}$ ions in $CaF_2$ can serve as an example of pair  REI  OCs. 
Let us consider such a crystal with concentration $c\sim 0.01$ of $Nd^{3+}$ centers and assume 
that the spectral width of laser $\Gamma_L \sim$ 0.1 GHz, inhomogeneous width of ZPL
$\Gamma_{inh} \sim$ THz  and the homogeneous width of ZPL is $\Gamma_h \lesssim \Gamma_L$.
In this case the concentration of OCs with the same (up to $\Gamma_L$) frequency  is
$c \Gamma_L/\Gamma_{inh} $  giving $R_0 \sim 100 a$ for the mean  size of the
microcrystal working  as a separate OQC instance.

Consider an ensemble of N=50 REIs closest to the excited one, which can act as an OQC instance. 
The mean size of this ensemble is $(N/c)^{1/3} a \sim 17 a$. 
According to \cite{ref-hizhnyakov21}, the strongest interaction between 
REIs in this ensemble of ions is the quadrupole-quadrupole, giving $\delta \sim $ GHz. 
This interaction exceeds $\Gamma_L$.
Therefore, in this case, fast CNOT gate operations can indeed be
successfully performed for all N=50 pairs working as qubits.

\section{Summary}

It fas found that optical centers of rare-earth ions (REIs) in fluoride crystals  can be used 
to create fast (with sampling time $10^{-9}\rm{s}$)  optical quantum computers (OQC). It was also 
found that pair optical centers in crystals with strong ZPL with large inhomogeneous and small 
homogeneous width in their spectrum are also promising systems for fast optical quantum computers: 
the dark states of the pair centers can be used as states of the qubits, while the bright states 
are suitable for CNOT and other conditional gate operations. A scheme of CNOT gate operation for 
pair centers is proposed, including the required sequence of laser pulses. 

Of particular interest for OQCs are  $SiV$ and $GeV$ pair optical centers in diamond, since these 
OCs can operate as qubits at elevated temperatures.
Pair optical centers of REIs in fluorite-type crystals are found to be also good candidates for OQCs;  
their use makes it possible to significantly expand the possibilities of choosing suitable REIs.


\begin{acknowledgments}
The work was supported by Estonian Research Council grant PRG347.
\end{acknowledgments}


\end{document}